\documentclass[linenumbers]{aastex631}

\usepackage{threeparttable}
\nolinenumbers
\shorttitle{The outburst of SLX 1746--331 in 2023}
\shortauthors{Peng et al.}
\begin{document}
\nolinenumbers
\title{Back to business: SLX 1746--331 after 13 years of silence}

\author[0000-0002-5554-1088]{Jing-Qiang Peng\textsuperscript{*}}
\email{pengjq@ihep.ac.cn}
\affiliation{Key Laboratory of Particle Astrophysics, Institute of High Energy Physics, Chinese Academy of Sciences, 100049, Beijing, China}
\affiliation{University of Chinese Academy of Sciences, Chinese Academy of Sciences, 100049, Beijing, China}
\author{Shu Zhang\textsuperscript{*}}
\email{szhang@ihep.ac.cn}
\affiliation{Key Laboratory of Particle Astrophysics, Institute of High Energy Physics, Chinese Academy of Sciences, 100049, Beijing, China}

\author[0000-0002-6454-9540]{Peng-Ju Wang\textsuperscript{*}}
\email{wangpj@ihep.ac.cn}
\affiliation{Institute f{\"u}r Astronomie und Astrophysik, Kepler Center for Astro and Particle Physics, Eberhard Karls, Universit{\"a}t, Sand 1, D-72076 T{\"u}bingen, Germany}
\author[0000-0001-5586-1017]{Shuang-Nan Zhang}
\affiliation{Key Laboratory of Particle Astrophysics, Institute of High Energy Physics, Chinese Academy of Sciences, 100049, Beijing, China}
\affiliation{University of Chinese Academy of Sciences, Chinese Academy of Sciences, 100049, Beijing, China}

\author[0000-0003-3188-9079]{Ling-Da Kong}
\affiliation{Institute f{\"u}r Astronomie und Astrophysik, Kepler Center for Astro and Particle Physics, Eberhard Karls, Universit{\"a}t, Sand 1, D-72076 T{\"u}bingen, Germany}

\author[0000-0001-8768-3294]{Yu-Peng Chen}
\affiliation{Key Laboratory of Particle Astrophysics, Institute of High Energy Physics, Chinese Academy of Sciences, 100049, Beijing, China}

\author[0000-0001-5160-3344]{Qing-Cang Shui}
\affiliation{Key Laboratory of Particle Astrophysics, Institute of High Energy Physics, Chinese Academy of Sciences, 100049, Beijing, China}
\affiliation{University of Chinese Academy of Sciences, Chinese Academy of Sciences, 100049, Beijing, China}

\author[0000-0001-9599-7285]{Long Ji}
\affiliation{School of Physics and Astronomy, Sun Yat-Sen University, Zhuhai, 519082, China}

\author[0000-0002-9796-2585]{Jin-Lu Qu}
\affiliation{Key Laboratory of Particle Astrophysics, Institute of High Energy Physics, Chinese Academy of Sciences, 100049, Beijing, China}

\author[0000-0002-2705-4338]{Lian Tao}
\affiliation{Key Laboratory of Particle Astrophysics, Institute of High Energy Physics, Chinese Academy of Sciences, 100049, Beijing, China}

\author[0000-0002-2749-6638]{Ming-Yu Ge}
\affiliation{Key Laboratory of Particle Astrophysics, Institute of High Energy Physics, Chinese Academy of Sciences, 100049, Beijing, China}
\author{Rui-Can MA}
\affiliation{Key Laboratory of Particle Astrophysics, Institute of High Energy Physics, Chinese Academy of Sciences, 100049, Beijing, China}
\affiliation{University of Chinese Academy of Sciences, Chinese Academy of Sciences, 100049, Beijing, China}
\author[0000-0003-4856-2275]{Zhi Chang}
\affiliation{Key Laboratory of Particle Astrophysics, Institute of High Energy Physics, Chinese Academy of Sciences, 100049, Beijing, China}

\author{Jian Li}
\affiliation{CAS Key Laboratory for Research in Galaxies and Cosmology, Department of Astronomy, University of Science and Technology of China, Hefei 230026, China}
\affiliation{School of Astronomy and Space Science, University of Science and Technology of China, Hefei 230026, China}
\author[0000-0003-2310-8105]{Zhao-sheng Li}
\affiliation{ Key Laboratory of Stars and Interstellar Medium, Xiangtan University, Xiangtan 411105, Hunan, China}

\author{Zhuo-Li Yu}
\affiliation{Key Laboratory of Particle Astrophysics, Institute of High Energy Physics, Chinese Academy of Sciences, 100049, Beijing, China}
\author{Zhe Yan}
\affiliation{University of Chinese Academy of Sciences, Chinese Academy of Sciences, 100049, Beijing, China}
\affiliation{Yunnan Observatories, Chinese Academy of Sciences, Kunming 650216, China}
\affiliation{Key Laboratory for the Structure and Evolution Celestial Objects, Chinese Academy of Sciences, Kunming 650216, China}
\affiliation{Center for Astronomical Mega-Science, Chinese Academy of Sciences, Beijing 100012, China}

\author{Peng Zhang}
\affiliation{College of Science, China Three Gorges University, Yichang 443002, China }
\affiliation{Center for Astronomy and Space Sciences, China Three Gorges University, Yichang 443002, China}

\author{Yun-Xiang Xiao}
\affiliation{Key Laboratory of Particle Astrophysics, Institute of High Energy Physics, Chinese Academy of Sciences, 100049, Beijing, China}
\affiliation{University of Chinese Academy of Sciences, Chinese Academy of Sciences, 100049, Beijing, China}

\author{Shu-Jie Zhao}
\affiliation{Key Laboratory of Particle Astrophysics, Institute of High Energy Physics, Chinese Academy of Sciences, 100049, Beijing, China}
\affiliation{University of Chinese Academy of Sciences, Chinese Academy of Sciences, 100049, Beijing, China}


\begin{abstract}
\nolinenumbers
The black hole candidate system SLX 1746--331 was back to business in 2023, after a long silence of roughly 13 years. An outburst was observed thoroughly by \textit{Insight}-HXMT and \textit{NICER}.  The outburst is characterized by spectral dominance of the soft state, where the joint \textit{Insight}-HXMT and \textit{NICER} spectral analysis shows the temperature dependence of the disk flux follows $T_{\rm in}^{3.98}$, and thus suggests that the inner disk reaches to ISCO during almost the entire outburst.  By assuming 0.3 $L_{\rm Edd}$ for the peak flux and an inclination angle of zero degrees, the lower limit of the compact object hosted in this system is estimated as 3.28$\pm 2.14 M_\odot$.  We also look into the relation of the disk temperature and disk flux for a sample of black hole systems, and by taking the disk temperature derived in the outburst of SLX 1746--331, such a relation results in a mass estimation of  $5.2 \pm 4.5M_\odot$.  Finally, the spin of the compact object is constrained to larger than 0.8 with a spectral model of kerrbb.

\end{abstract}



\section{Introduction} 
\label{sec:intro}

Low-mass X-ray binaries (LMXRBs) consist of a star and a compact star, which is typically a black hole (BH) or a Neutron Star (NS). The typical mass of the companion star in LMXRBs is relatively small ($\lesssim 1M_\odot$). The compact object accretes material from the companion star via the Roche-lobe \citep{1973Shakura}.
Some black hole X-ray binaries are known as persistent sources, as they maintain relatively high accretion rates for long periods \citep{2009Deegan}. In contrast, transient sources remain in a quiescent state with a low accretion rate for prolonged periods and are generally challenging to be observed in the X-ray band. When they enter the outburst state, their X-ray emission intensifies and can last from a few days to several months \citep{1995Cannizzo, 2001Lasota,2011Belloni,2016Corral-Santana}.

For the outbursts of classical black hole X-ray transient sources, their trajectory on the Hardness-Intensity diagram (HID) is so-called  "q" shape \citep{2001Homan, 2004Fender, 2012Motta}.
Low-mass black hole X-ray binaries can be classified into different spectral states during their outburst cycles, including the Low/Hard States (LHS), High/Soft States (HSS), and Intermediate States (IMS)  \citep{2005Belloni,2009Motta}.
In the LHS, the emission is primarily from corona/jet hard X-ray emission, with the non-thermal component dominating and the photon index $\Gamma$ typically around 1.5. The accretion disk is generally believed to be truncated, with a truncation radius typically greater than 100 $R_{\rm g}$. However, there are some outbursts in which the accretion disk is not truncated in the LHS \citep{2006Miller,2010Done}.
As the luminosity increases, the radius of the disk moves inward, the temperature of the inner disk rises \citep{2008Gierlinski}, and the disk reaches its innermost stable circular orbit (ISCO) as it enters into the HSS. At this time, the thermal emission from the disk dominates, and the corona/jet becomes almost invisible with a photon index of about 2.1-3.7 \citep{2006McClintock, 1997Esin}.
The IMS can be further divided into the hard intermediate state (HIMS) and the soft intermediate state (SIMS), where the energy spectrum of SIMS is softer than that of HIMS \citep{2005Homan}.

In a classical successful outburst, a low-mass black hole X-ray binary transfers from the quiescent state to the LHS and then through the IMS to the HSS and back to the LHS as the accretion rate increases, following a counterclockwise "q" shape on the HID \citep{2001Homan,2005Homan,2016Belloni}.
However, not all outbursts follow a complete "q" shape. In some sources, the accretion rate is so low that the outburst does not reach the HSS but only evolves to the low/hard or intermediate state before ending. These outbursts are known as "failed outbursts" and account for approximately 38$\%$ of all outbursts \citep{2004Brocksopp,2009Capitanio,2016Tetarenko}.
In contrast to the absence of LHS, there is a class of outbursts that do not have a LHS at the beginning of the outburst, or whose LHS is not observed. Such outbursts are relatively rare, as seen in sources like 4U 1630--472 and MAXI J0637--430 \citep{2020Baby,2022Ma}.

SLX 1746--331 is a transient low-mass X-ray binary located at the galactic center, and was identified within the surveyed fields of the Einstein galactic plane survey conducted by \cite{1984Hertz}. It  was discovered with the Spacelab 2 XRT in 1985 August and detected by the RASS in 1990 \citep{1988Warwick,1990Skinner,1998Motch,2003Wilson}.
SLX 1746--331 had a very soft spectrum, which is best fitted with thermal bremsstrahlung with a temperature of $kT$=1.5 keV.
\citep{1990Skinner} suggested that SLX 1746--331 may be a potential black hole candidate based on its very soft spectrum.
In history, outburst of SLX 1746--331 were observed by \textit{RXTE/PCA} \citep{2007Markwardt} and \textit{INTEGRAL/JEM-X} in 2007/2008 \citep{2008Kuulkers}  and by \textit{MAXI} in 2010 \citep{2011Ozawa}. Both outbursts exhibited a very soft spectrum and are thought to be associated with black hole candidates (BHCs) in a soft state.
\cite{2015Yan} estimated the distance of SLX 1746--331 to be about 10.81 $\pm$ 3.52 kpc using data from \textit{Rossi X-ray Timing Explorer (RXTE).}

SLX 1746--331 underwent outburst again in 2023 after the long silence of roughly 13 years, and was observed thoroughly by \textit{Insight}-HXMT and \textit{NICER}. SLX 1746--331 was observed a total of 46 times by \textit{Insight}-HXMT over 35 days, from March 14th to April 17th, 2023. And \textit{NICER} conducted 48 observations of SLX 1746--331 over 93 days, from March 8th to June 8th. In this letter, we perform a detailed spectral analysis and reveal some properties regarding both the outburst and the harbored compact object. In Section \ref{obser}, we describe the observations and data reduction. The detailed results are presented in Section \ref{result}. The results are discussed and the conclusions are presented in Section \ref{dis}.

\section{Observations and Data reduction}
\label{obser}

\begin{table}[]
    \centering
		\caption{ \textit{Insight}-HXMT and \textit{NICER} observations of SLX 1746--331 during the 2023 outburst. }
		\begin{tabular}{ccccccc}
		  \hline
              \hline
        \textit{Insight}-HXMT &  Observed date & LE exposure time & ME exposure time&\textit{NICER} & Observed date & Exposure Time 
         \\ ObsID& (MJD) & (s) & (s) &ObsID &(MJD)&(s)
       \\ \hline
       P051436300101 & 60017.14  &1821 & 2703& 6203700101 &60011.69 & 1462 \\ 
       P051436300102 & 60017.29 &260.3 & 1707& 6203700102 &60012.02 & 1932  \\ 
       P051436300103 & 60017.43 &270.3 & 922.5&6203700103 &60013.06 & 3014 \\ 
       P051436300201 & 60018.53 &1614 & 2297&6203700104 &60014.21 & 839  \\ 
       P051436300202 & 60018.71 &1822 & 2947&6203700105 &60015.05 & 766  \\ 
       P051436300301 & 60020.38 &352.1 & 1949&6203700108 &60023.00 & 1306 \\ 
       P051436300302 & 60020.56 &2633 & 2678&6203700109 &60024.14 & 2872    \\ 
       P051436300501 & 60021.38 & 817 & 1973&6203700110 &60025.30 & 3888   \\ 
       P051436300502 & 60021.55 & 2574 & 2664&6203700111 &60026.08 & 2212   \\ 
       P051436300601 & 60022.57 & 478.8 & 2178&6203700112 &60027.26 & 724    \\ 
       P051436300602 & 60022.74 & 1077 & 2018& 6203700113 &60028.46 & 1838   \\ 
       P051436300603 & 60022.88 & 359.1 & 1079&6203700114 &60029.56 & 2101  \\   
       P051436300701 & 60023.56 & 1937 & 2650 &6203700115 &60031.24 & 1598 \\  
       P051436300702 & 60023.74 & 1137 & 2020&6203700116 &60032.67 & 1584   \\ 
       P051436300703 & 60023.87& 538.7 & 1066 &6203700117 &60039.33 & 1700   \\
       P051436300801 & 60024.95& 1763 & 2165& 6203700118 &60043.01 & 2079   \\
       P051436300802 & 60025.08& 955.6 & 1658& 6203700119 &60044.03 & 1831  \\
       P051436300803 & 60025.22& 284.3 & 964.2&6203700120 &60046.73 & 1552    \\
       P051436300901 & 60026.93& 2723 & 3235 &6203700121 &60047.57 & 1200 \\
       P051436301001 & 60028.19& 627.4 & 798.9&6203700122 &60051.66 & 1797   \\
       P051436301002 & 60028.34& 886.8 & 2061&6203700123 &60052.68 & 1846   \\
       P051436301003 & 60028.50& 119.7 & 657.4& 6203700124 &60053.25 & 1675  \\
       P051436301101 & 60030.04& 179.5 & 1445 &6203700125 &60054.69 & 1742  \\
       P051436301102 & 60030.18& 579.5 & 844.2& 6203700126 &60055.84 & 1312  \\
       P051436301103 & 60030.32& 716.2 & 1142& 6203700127 &60056.49 & 2483   \\
       P051436301301 & 60032.75& 718.2 & 3469 &6203700128 &60060.68 & 2297 \\
       P051436301302 & 60032.93& 1886 & 1554&6203700129 &60061.65 & 1227  \\
       P051436301303 & 60033.06& 299.2 & 824& 6203700130 &60062.48 & 240  \\
       P051436301401 & 60034.54& 1495 & 3029 &6203700131 &60063.45 & 724 \\
       P051436301402 & 60034.72& 210.5 & 3814&6203700132 &60064.35 & 945   \\
       P051436301601 & 60036.32& 1915 & 5360& 6203700133 &60065.00 & 275 \\
       P051436301801 & 60040.03& 1205 & 3301&6203700134 &60074.12 & 80  \\
       P051436301802 & 60040.14& 770.1 & 7371&6203700135 &60080.30 & 911  \\    
       P051436301901 & 60041.55& 1927 & 5947& 6203700136 &60081.39 & 1993 \\ 
       P051436302001 & 60043.20& 3522 & 4340&6203700137 &60082.82 & 711   \\ 
       P051436302002 & 60043.44& 880.8 & 1040&6203700138 &60083.27 & 1129   \\ 
       P051436302101 & 60046.04& 2499 & 2181 & 6203700139 &60084.50 & 1557\\ 
       P051436302102 & 60046.17& 1436& 1972&6203700140 &60085.14 & 1503 \\ 
       P051436302103 & 60046.30& 1058& 1227&6203700141 &60086.24 & 613 \\ 
       P051436302201 & 60047.83& 3711& 3709& 6203700142 &60087.28 & 52.58  \\ 
       P051436302202 & 60048.19& 3305& 3332&6203700143 &60088.32 & 23 \\ 
       P051436302301 & 60050.08& 2503& 3064&6203700144 &60089.02 & 770 \\ 
       P051436302302 & 60050.27& 777.1& 842.1& 6203700145 &60089.10 & 1467 \\ 
       P051436302401 & 60051.14& 2454& 2565 &6203700146 &60093.52 & 1683\\ 
       P051436302403 & 60051.43& 1137& 996.1&6203700147 &60094.76 & 327  \\ 
       P051436302404 & 60051.57& 2334& 2469 &6203700148 &60096.23 & 720\\   
       &&&& 6203700149 &60102.81 & 1410    \\      
        &&&&6203700150 &60103.64 & 1152   \\ 
    \hline
        \label{observation}    

    \end{tabular}

\end{table}

\subsection{\textit{Insight}-HXMT}
\textit{Insight}-HXMT is the first Chinese X-ray astronomy satellite, which was successfully launched on 2017 June 15 \citep{2014Zhang, 2018Zhang, 2020Zhang}.  It carries three scientific payloads: the low energy X-ray telescope (LE, SCD detector, 1--15 keV, 384 $\rm cm^{2}$,
\citealt{2020Chen}), the medium energy X-ray telescope (ME, Si-PIN detector, 5--35 keV, 952 $\rm cm^{2}$, \citealt{2020Cao} ), and the high energy X-ray telescope (HE, phoswich NaI(CsI), 20--250 keV, 5100 $\rm cm^{2}$, \citealt{2020Liu}).

\textit{Insight}-HXMT started to observe SLX 1746--331 at the peak of its outburst on March 14, 2023 and continued until May 19, 2023.
We extract the data from LE and ME using the \textit{Insight}-HXMT Data Analysis software {\tt{HXMTDAS v2.05}}. 
The data are filtered with the  criteria recommended of the \textit{Insight}-HXMT Data Reduction Guide {\tt v2.05}\footnote[1]{{http://hxmtweb.ihep.ac.cn/SoftDoc/648.jhtml}}
{\tt Xspec v12.13.0c}\footnote[2]{{https://heasarc.gsfc.nasa.gov/docs/xanadu/xspec/index.html}} is used to perform analysis of spectrum, where  LE data above 2 keV are chosen to  suppress possible source pollution at lower energies.
Due to the count rate of high-energy photons in SLX 1746--331, the energy bands considered for spectral analysis are LE 2--8 keV and ME 8--25 keV. One percent  systematic error is added to data  \citep{2020Liao}, and errors are estimated via  Markov Chain Monte-Carlo (MCMC) chains with a length of 20000.

\subsection{\textit{NICER}}
The X-ray TIming Instrument (XTI) of the Neutron Star Interior Composition Explorer (\textit{NICER}) is an International Space Station (ISS) payload, which was launched by the Space X Falcon 9 rocket on 3 June 2017 \citep{2016Gendreau}. And \textit{NICER} has a large effective area and high temporal resolution in soft X-ray band (0.2--12 keV), which  may well fit the black body component at low temperatures.

\textit{NICER} started to observe SLX 1746--331 on March 8, 2023 and continued until June 8, 2023. The \textit{NICER} observations covered the rising phase of the SLX 1746--331 outburst.
\textit{NICER} data are reduced using the standard pipeline tool {\tt \textit{NICER}l2\footnote[3]{https://heasarc.gsfc.nasa.gov/lheasoft/ftools/headas/\textit{NICER}l2.html}}. 
We used "overonly\underline{ }range" and "underonly\underline{ }range" in the \textit{NICER}l2 to filter the data. This helped ensure that overshoots (0--1) and undershoots (0--200) were within the recommended range for analysis.
We extract light curves using {\tt \textit{NICER}l3-lc\footnote[4]{https://heasarc.gsfc.nasa.gov/docs/software/lheasoft/ftools/headas/\textit{NICER}l3-lc.html}} in 1--4 keV, 4--10 keV and 1--10 keV.
We use {\tt \textit{NICER}l3-spect\footnote[5]{https://heasarc.gsfc.nasa.gov/docs/software/lheasoft/help/\textit{NICER}l3-spect.html}} to extract the spectrum, with "{\tt nibackgen3C50\footnote[6]{https://heasarc.gsfc.nasa.gov/docs/\textit{NICER}/analysis\_{}threads/background/}}" model to estimate the
background for the spectral analysis.
For the fitting of the spectrum, we choose an energy range of 0.5--10 keV.

\section{Results}
\label{result}

\subsection{Light curve and Hardness-intensity diagram}
\label{light curve}

\begin{figure*}
	\centering
	\includegraphics[angle=0,scale=0.35]{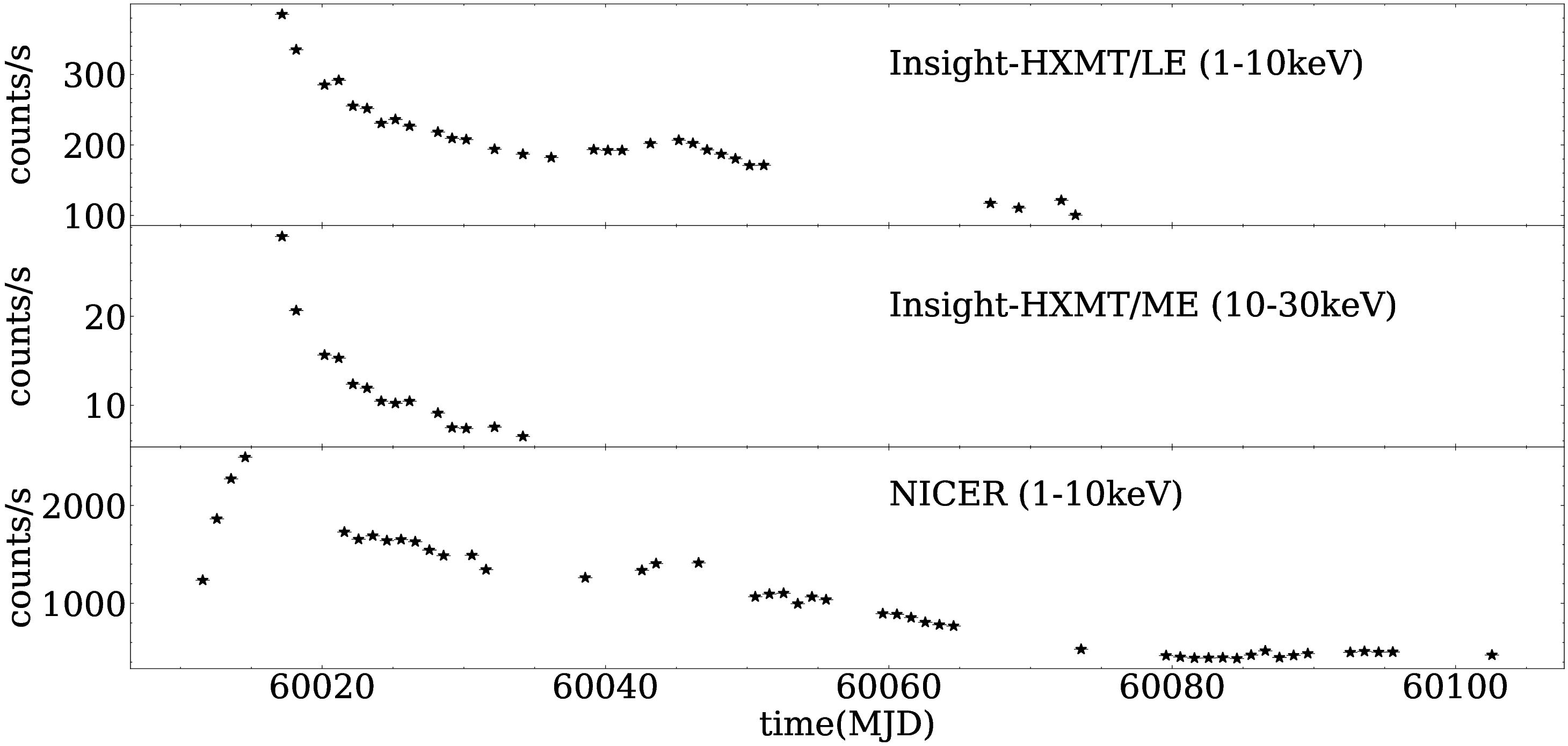}
	\caption{Light curve of SLX 1764--331 during the 2023 outburst.
Top panel: the light curve of \textit{Insight}-HXMT LE 1--10 keV
Middle panel: the light curve of \textit{Insight}-HXMT ME 10--30 keV
Bottom panel: the light curve for \textit{NICER} 1--10 keV.}
	\label{lcurve}
\end{figure*}

\begin{figure}
	\centering
	\includegraphics[angle=0,scale=0.31]{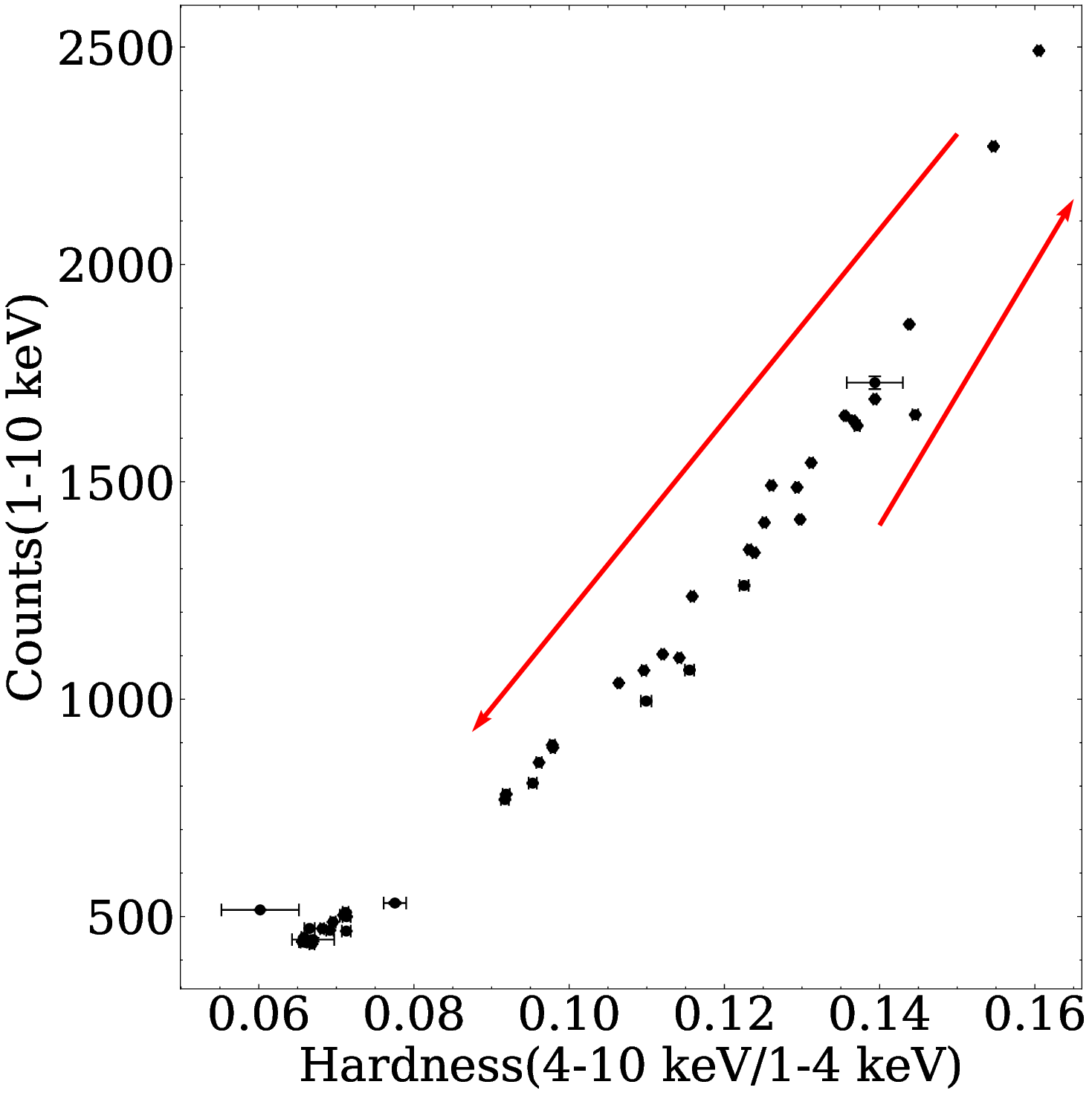}
	\caption{The \textit{NICER} hardness-intensity diagram of SLX 1746--331, where the hardness is defined as the ratio of 4--10 keV to 1--4 keV count rate. The red arrows represent the direction of the evolution.}
	\label{HID}
\end{figure}

As shown in Figure \ref{lcurve}, since the alert from \textit{MAXI} for the outburst of SLX 1746--331, \textit{NICER} observed from March 8 until June 8, covering both the rise and decay phases of the outburst. \textit{Insight}-HXMT observed from the outburst peak on March 14 to the decay phase till May 19 (Table \ref{observation}).

As \textit{NICER} observations cover the rise and decay of the outburst of SLX 1746--331, we extract the 1--4 keV, 4-10 keV, and 1--10 keV light curves of \textit{NICER} to construct the Hardness-Intensity Diagram (HID) of SLX 1746--331. Figure \ref{HID} shows the \textit{NICER} HID of SLX 1746--331 with hardness defined as the count rate ratio of 4--10 keV to 1--4 keV.
The red arrows in the HID represent the trajectory of SLX 1746--331, where it is observed that the outburst of SLX 1746--331 begins in a soft state, then slightly hardens, and eventually evolves toward softer.

We use the tool {\tt{powspec}}\footnote[7]{{https://heasarc.gsfc.nasa.gov/docs/xanadu/xronos/examples/powspec.html}} of {\tt{HEASOFT}}  to generate power spectra with  a time resolution of 0.008 s, the
interval of 65.5 s and Nyquist frequency of 62.5 Hz, under the Miyamoto normalization \citep{1990Belloni, 1992Miyamoto}.
Only a very weak power-law noise is visible in the power density spectrum (PDS), and no QPO is detected.
These indicate that it remains in a soft state throughout the outburst.

\subsection{Evolution of the Spectral parameters}
\label{parameters}

\begin{table}[]
    \centering
		\caption{Fitting results of \textit{NICER} for Model M1. $N_{\rm H}$ is the interstellar absorption, $T_{\rm in}$ is the inner disk temperature, Norm(disk) is the normalization of the disk, and Flux is the flux of the unabsorbed disk from 1 to 100 keV.}
		\begin{tabular}{cccccc}
		 \hline
		   \hline
          \textit{NICER} & $N_{\rm H}$ & $T_{\rm in}$ &Norm(disk)&Flux& $\chi^2$/dof
         \\ObsID &$10^{22}$ $\rm cm^{-2}$&(keV)&&$10^{-8}$~erg~s$^{-1}$~cm$^{-2}$
       \\ \hline
        
        6203700101 &$0.63_{-0.004}^{+0.005}$ & $1.51_{-0.01}^{+0.01}$&$70.8_{-1.3}^{+1.4}$&$0.65_{-0.002}^{+0.002}$ &279.96/142\\ 
        6203700102 &$0.65_{-0.003}^{+0.003}$ & $1.70_{-0.01}^{+0.01}$&$67.3_{-1.1}^{+1.1}$&$1.04_{-0.003}^{+0.004}$&172.11/148 \\ 
        6203700103 &$0.66_{-0.002}^{+0.003}$ & $1.79_{-0.01}^{+0.01}$&$70.6_{-1.0}^{+1.0}$&$1.38_{-0.002}^{+0.006}$ &229.20/153    \\ 
        6203700104 &$0.66_{-0.003}^{+0.003}$ & $1.92_{-0.01}^{+0.01}$&$68.5_{-1.2}^{+1.2}$&$1.79_{-0.011}^{+0.004}$&170.92/146  \\ 
        6203700105 &$0.66_{-0.003}^{+0.003}$ & $1.91_{-0.01}^{+0.01}$&$69.9_{-1.2}^{+1.2}$&$1.77_{-0.002}^{+0.011}$& 170.09/147   \\ 
        6203700108 &$0.64_{-0.003}^{+0.003}$ & $1.77_{-0.01}^{+0.01}$&$69.0_{-1.0}^{+1.1}$&$1.09_{-0.003}^{+0.005}$&189.56/149  \\ 
        6203700109 &$0.65_{-0.003}^{+0.003}$ & $1.70_{-0.01}^{+0.01}$&$67.5_{-1.1}^{+1.0}$&$1.05_{-0.005}^{+0.002}$&172.98/152 \\ 
        6203700110 &$0.65_{-0.004}^{+0.004}$ & $1.68_{-0.01}^{+0.01}$&$68.9_{-1.0}^{+1.0}$&$1.02_{-0.001}^{+0.006}$&172.48/152  \\ 
        6203700111 &$0.65_{-0.003}^{+0.003}$ & $1.67_{-0.01}^{+0.01}$&$68.3_{-1.1}^{+1.1}$&$0.99_{-0.009}^{+0.005}$&202.59/152   \\ 
        6203700112 &$0.65_{-0.004}^{+0.004}$ & $1.68_{-0.01}^{+0.01}$&$67.8_{-1.3}^{+1.4}$&$1.02_{-0.011}^{+0.009}$&138.41/143  \\ 
        6203700113 &$0.65_{-0.003}^{+0.003}$ & $1.64_{-0.01}^{+0.01}$&$69.1_{-1.1}^{+1.0}$&$0.92_{-0.004}^{+0.002}$ &165.74/151   \\ 
        6203700114 &$0.65_{-0.004}^{+0.004}$ & $1.62_{-0.01}^{+0.01}$&$69.7_{-1.1}^{+1.1}$&$0.90_{-0.005}^{+0.001}$ &184.06/151   \\      
        6203700115 &$0.65_{-0.004}^{+0.004}$ & $1.60_{-0.01}^{+0.01}$&$71.6_{-1.2}^{+1.2}$&$0.85_{-0.002}^{+0.007}$&168.58/149  \\  
        6203700116 &$0.64_{-0.004}^{+0.004}$ & $1.57_{-0.01}^{+0.01}$&$70.4_{-1.2}^{+1.2}$&$0.79_{-0.006}^{+0.003}$&167.93/149  \\ 
        6203700117 &$0.63_{-0.005}^{+0.005}$ & $1.58_{-0.01}^{+0.01}$&$65.1_{-1.5}^{+1.5}$&$0.75_{-0.003}^{+0.004}$ &131.17/135   \\ 
        6203700118 &$0.64_{-0.003}^{+0.003}$ & $1.60_{-0.01}^{+0.01}$&$65.9_{-1.0}^{+1.0}$&$0.77_{-0.002}^{+0.006}$ &172.50/150  \\ 
        6203700119 &$0.64_{-0.003}^{+0.003}$ & $1.59_{-0.01}^{+0.01}$&$67.4_{-1.1}^{+1.1}$&$0.80_{-0.004}^{+0.004}$ &177.61/150   \\ 
        6203700120 &$0.64_{-0.003}^{+0.003}$ & $1.57_{-0.01}^{+0.01}$&$72.5_{-1.2}^{+1.2}$&$0.81_{-0.002}^{+0.006}$ &178.51/152   \\      
        6203700121 &$0.64_{-0.003}^{+0.004}$ & $1.56_{-0.01}^{+0.01}$&$69.2_{-1.2}^{+1.2}$&$0.74_{-0.003}^{+0.004}$&148.69/145  \\ 
        6203700122 &$0.63_{-0.003}^{+0.004}$ & $1.51_{-0.01}^{+0.01}$&$66.6_{-1.1}^{+1.1}$&$0.63_{-0.005}^{+0.003}$ &158.54/145\\ 
        6203700123 &$0.63_{-0.003}^{+0.004}$ & $1.49_{-0.01}^{+0.01}$&$67.3_{-1.1}^{+1.1}$&$0.61_{-0.005}^{+0.003}$&153.22/145 \\ 
        6203700124 &$0.64_{-0.003}^{+0.004}$ & $1.48_{-0.01}^{+0.01}$&$68.9_{-1.1}^{+1.1}$&$0.60_{-0.003}^{+0.002}$ &164.67/145    \\ 
        6203700125 &$0.64_{-0.003}^{+0.003}$ & $1.45_{-0.01}^{+0.01}$&$70.0_{-1.2}^{+1.2}$&$0.57_{-0.002}^{+0.003}$&151.27/145  \\ 
        6203700126 &$0.64_{-0.003}^{+0.003}$ & $1.44_{-0.01}^{+0.01}$&$69.9_{-1.2}^{+1.2}$&$0.54_{-0.002}^{+0.005}$&172.30/141   \\ 
        6203700127 &$0.64_{-0.003}^{+0.003}$ & $1.43_{-0.01}^{+0.01}$&$71.8_{-1.1}^{+1.1}$&$0.54_{-0.002}^{+0.002}$&213.12/180  \\ 
        6203700128 &$0.64_{-0.003}^{+0.003}$ & $1.36_{-0.01}^{+0.01}$&$73.8_{-1.2}^{+1.2}$&$0.50_{-0.006}^{+0.012}$&177.31/144 \\ 
        6203700129 &$0.64_{-0.004}^{+0.004}$ & $1.34_{-0.01}^{+0.01}$&$74.0_{-1.4}^{+1.4}$&$0.43_{-0.001}^{+0.002}$ &156.15/140  \\ 
        6203700130 &$0.63_{-0.006}^{+0.006}$ & $1.35_{-0.01}^{+0.01}$&$71.3_{-2.3}^{+2.4}$&$0.42_{-0.004}^{+0.002}$ &154.03/125   \\ 
        6203700131 &$0.64_{-0.005}^{+0.005}$ & $1.34_{-0.01}^{+0.01}$&$73.1_{-1.8}^{+1.8}$&$0.41_{-0.005}^{+0.003}$ &177.01/134  \\ 
        6203700132 &$0.63_{-0.004}^{+0.004}$ & $1.31_{-0.01}^{+0.01}$&$73.6_{-1.5}^{+1.6}$&$0.39_{-0.001}^{+0.002}$&166.93/136   \\ 
        6203700133 &$0.63_{-0.006}^{+0.006}$ & $1.31_{-0.01}^{+0.01}$&$73.7_{-2.3}^{+2.4}$&$0.38_{-0.001}^{+0.003}$ &139.89/125   \\      
        6203700134 &$0.63_{-0.005}^{+0.005}$ & $1.17_{-0.02}^{+0.02}$&$73.3_{-2.0}^{+2.3}$&$0.26_{-0.003}^{+0.004}$&139.03/103  \\  
        6203700135 &$0.62_{-0.005}^{+0.005}$ & $1.13_{-0.01}^{+0.01}$&$76.5_{-1.9}^{+2.0}$&$0.21_{-0.001}^{+0.002}$&189.92/131  \\ 
        6203700136 &$0.62_{-0.005}^{+0.005}$ & $1.12_{-0.01}^{+0.01}$&$76.5_{-1.9}^{+2.0}$&$0.20_{-0.002}^{+0.002}$&171.71/136   \\ 
        6203700137 &$0.63_{-0.004}^{+0.004}$ & $1.11_{-0.01}^{+0.01}$&$77.1_{-1.6}^{+1.6}$&$0.19_{-0.001}^{+0.001}$&151.01/126  \\ 
        6203700138 &$0.63_{-0.005}^{+0.005}$ & $1.10_{-0.01}^{+0.01}$&$79.4_{-2.2}^{+2.3}$&$0.19_{-0.002}^{+0.003}$&146.92/131   \\ 
        6203700139 &$0.62_{-0.005}^{+0.005}$ & $1.10_{-0.01}^{+0.01}$&$77.8_{-1.9}^{+2.0}$&$0.19_{-0.002}^{+0.002}$&172.00/135   \\      
        6203700140 &$0.62_{-0.004}^{+0.005}$ & $1.12_{-0.01}^{+0.01}$&$73.6_{-1.7}^{+1.7}$&$0.19_{-0.001}^{+0.002}$&150.89/134  \\ 
        6203700141 &$0.62_{-0.005}^{+0.005}$ & $1.12_{-0.01}^{+0.01}$&$74.5_{-1.7}^{+1.8}$&$0.19_{-0.001}^{+0.002}$ &127.04/109  \\    
        6203700144 &$0.63_{-0.005}^{+0.005}$ & $1.13_{-0.01}^{+0.01}$&$78.2_{-2.0}^{+2.1}$&$0.21_{-0.002}^{+0.002}$&142.64/127  \\  
        6203700145 &$0.63_{-0.004}^{+0.004}$ & $1.14_{-0.01}^{+0.01}$&$76.7_{-1.7}^{+1.7}$&$0.21_{-0.002}^{+0.002}$ &154.44/137  \\ 
        6203700146 &$0.63_{-0.004}^{+0.004}$ & $1.15_{-0.01}^{+0.01}$&$76.2_{-1.6}^{+1.6}$&$0.23_{-0.001}^{+0.001}$ &158.31/137   \\ 
        6203700147 &$0.63_{-0.007}^{+0.007}$ & $1.14_{-0.01}^{+0.01}$&$79.82_{-2.8}^{+2.9}$&$0.23_{-0.002}^{+0.004}$&112.28/120  \\ 
        6203700148 &$0.63_{-0.005}^{+0.005}$ & $1.15_{-0.01}^{+0.01}$&$78.46_{-2.1}^{+2.2}$&$0.23_{-0.002}^{+0.003}$&157.26/128   \\ 
        6203700149 &$0.62_{-0.005}^{+0.005}$ & $1.13_{-0.01}^{+0.01}$&$76.7_{-1.7}^{+1.8}$&$0.21_{-0.001}^{+0.002}$ &171.09/134   \\      
        6203700150 &$0.62_{-0.005}^{+0.005}$ & $1.13_{-0.01}^{+0.01}$&$75.6_{-1.8}^{+1.9}$&$0.20_{-0.001}^{+0.002}$ &161.08/132  \\ 
        
    \hline
        \label{nicerpara} &     

    \end{tabular}

\end{table}

\begin{table}[]
    \centering
		\caption{Fitting results of \textit{Insight}-HXMT for Model M2. $T_{\rm in}$ is the inner disk temperature, Norm(disk) is the normalisation of the disk, $\Gamma$ is the photon index, and Norm(powerlaw) is the normalisation of the powerlaw component. }
		\begin{tabular}{cccccc}
		  \hline
		   \hline
          \textit{Insight}-HXMT & $T_{\rm in}$ &Norm(disk)& $\Gamma$ &Norm(powerlaw)&$\chi^2$/dof
         \\ObsID &(keV)&
       \\ \hline
       P051436300101 & $1.78_{-0.01}^{+0.01}$  &$83.5_{-3.0}^{+2.88}$ & $2.09_{-0.10}^{+0.08}$&$0.51_{-0.13}^{+0.13}$&256.11/323 \\ 
       P051436300102 &  $1.86_{-0.02}^{+0.02}$  &$67.2_{-3.80}^{+2.10}$ & $2.10_{-0.13}^{+0.12}$&$0.52_{-0.16}^{+0.19}$&253.43/323   \\ 
       P051436300103 &  $1.78_{-0.02}^{+0.02}$  &$82.9_{-4.7}^{+3.8}$ & $2.03_{-0.27}^{+0.17}$&$0.35_{-0.20}^{+0.22}$&309.36/323  \\ 
       P051436300201 &  $1.75_{-0.01}^{+0.01}$  &$81.2_{-2.7}^{+2.7}$ & $2.13_{-0.13}^{+0.10}$&$0.37_{-0.12}^{+0.12}$&238.04/323  \\ 
       P051436300202 &  $1.72_{-0.01}^{+0.01}$  &$80.9_{-2.5}^{+2.2}$ & $2.02_{-0.11}^{+0.12}$&$0.27_{-0.07}^{+0.10}$&236.79/323   \\ 
       P051436300301 &  $1.75_{-0.02}^{+0.02}$  &$67.6_{-4.7}^{+4.5}$ & $2.28_{-0.27}^{+0.18}$&$0.39_{-0.20}^{+0.24}$&273.27/323   \\ 
       P051436300302 & $1.68_{-0.01}^{+0.01}$  &$78.3_{-2.7}^{+2.2}$ & $1.96_{-0.15}^{+0.14}$&$0.19_{-0.06}^{+0.09}$&247.27/323     \\ 
       P051436300501 &$1.73_{-0.02}^{+0.03}$  &$70.3_{-4.7}^{+3.8}$ & $2.18_{-0.17}^{+0.18}$&$0.35_{-0.13}^{+0.23}$&251.18/323   \\ 
       P051436300502 &$1.67_{-0.01}^{+0.01}$  &$80.3_{-2.8}^{+2.1}$ & $1.95_{-0.17}^{+0.14}$&$0.19_{-0.08}^{+0.10}$&259.19/323   \\ 
       P051436300601 & $1.67_{-0.02}^{+0.02}$  &$73.4_{-5.1}^{+3.3}$ & $2.07_{-0.20}^{+0.26}$&$0.23_{-0.10}^{+0.24}$&245.89/323    \\ 
       P051436300602 & $1.64_{-0.01}^{+0.01}$  &$78.4_{-3.5}^{+2.5}$ & $1.95_{-0.15}^{+0.21}$&$0.15_{-0.05}^{+0.12}$&263.53/323   \\ 
       P051436300603 & $1.66_{-0.02}^{+0.02}$  &$69.6_{-4.5}^{+5.4}$ & $2.39_{-0.20}^{+0.15}$&$0.44_{-0.20}^{+0.20}$&278.92/323  \\ 
       P051436300701 &$1.62_{-0.01}^{+0.01}$  &$77.5_{-3.5}^{+2.2}$ & $1.86_{-0.17}^{+0.25}$&$0.22_{-0.05}^{+0.14}$&251.14/323   \\  
       P051436300702 &$1.64_{-0.01}^{+0.02}$  &$73.7_{-4.4}^{+3.7}$ & $2.25_{-0.19}^{+0.19}$&$0.29_{-0.12}^{+0.18}$&243.51/323    \\ 
       P051436300703 &$1.63_{-0.01}^{+0.02}$  &$76.3_{-4.7}^{+3.7}$ & $1.96_{-0.35}^{+0.30}$&$0.15_{-0.09}^{+0.20}$&256.22/323   \\
       P051436300801 &$1.58_{-0.01}^{+0.01}$  &$84.6_{-4.1}^{+2.2}$ & $2.14_{-0.12}^{+0.12}$&$0.21_{-0.06}^{+0.16}$&243.52/323  \\
       P051436300802 &$1.61_{-0.01}^{+0.02}$  &$75.9_{-4.5}^{+4.2}$ & $2.28_{-0.20}^{+0.15}$&$0.34_{-0.14}^{+0.17}$&256.91/323   \\
       P051436300803 &$1.60_{-0.02}^{+0.03}$  &$79.2_{-7.68}^{+4.6}$ & $2.27_{-0.22}^{+0.35}$&$0.23_{-0.11}^{+0.33}$&242.56/323   \\
       P051436300901 &$1.55_{-0.01}^{+0.01}$  &$86.7_{-3.7}^{+2.6}$ & $2.20_{-0.12}^{+0.13}$&$0.30_{-0.08}^{+0.13}$&243.46/323  \\
       P051436301001 &$1.50_{-0.01}^{+0.02}$  &$96.5_{-6.7}^{+3.9}$ & $2.04_{-0.23}^{+0.30}$&$0.16_{-0.08}^{+0.20}$&283.36/323  \\
       P051436301002 &$1.69_{-0.02}^{+0.03}$  &$54.3_{-4.9}^{+4.5}$ & $2.47_{-0.15}^{+0.14}$&$0.64_{-0.20}^{+0.26}$&284.50/323   \\
       P051436301003 &$1.54_{-0.02}^{+0.04}$  &$81.1_{-6.1}^{+4.4}$ & $2.00_{-0.30}^{+0.46}$&$0.16_{-0.09}^{+0.44}$&251.22/323   \\
     P051436301101
     &$1.57_{-0.02}^{+0.04}$  &$71.6_{-8.5}^{+5.39}$&$2.36_{-0.18}^{+0.26}$&$0.38_{-0.15}^{+0.35}$&268.59/323    \\
       P051436301102 &$1.52_{-0.02}^{+0.01}$  &$86.9_{-3.80}^{+3.70}$&$2.29_{-0.22}^{+0.15}$&$0.26_{-0.01}^{+0.08}$&228.04/323   \\
       P051436301103 &$1.53_{-0.02}^{+0.02}$  &$79.9_{-6.31}^{+6.12}$&$2.17_{-0.37}^{+0.23}$&$0.23_{-0.15}^{+0.20}$&231.82/323  \\
       P051436301301 &$1.51_{-0.01}^{+0.01}$  &$84.1_{-3.60}^{+2.61}$&$1.67_{-0.21}^{+0.26}$&$0.16_{-0.03}^{+0.07}$&263.65/323 \\
       P051436301302 &$1.50_{-0.01}^{+0.01}$  &$83.18_{-4.35}^{+3.86}$&$2.01_{-0.25}^{+0.23}$&$0.15_{-0.07}^{+0.14}$&218.54/323 \\
        P051436301303
        &$1.47_{-0.01}^{+0.02}$  &$90.71_{-7.15}^{+4.55}$&$1.87_{-0.30}^{+0.30}$&$0.12_{-0.07}^{+0.16}$&266.04/323  \\
       P051436301401 &$1.51_{-0.01}^{+0.01}$  &$76.5_{-3.8}^{+3.5}$&$2.07_{-0.32}^{+0.20}$&$0.15_{-0.07}^{+0.15}$&268.30/323  \\
       P051436301402 &$1.49_{-0.02}^{+0.02}$  &$81.1_{-5.2}^{+5.6}$&$2.06_{-0.25}^{+0.24}$&$0.15_{-0.08}^{+0.12}$&287.45/323  \\
       P051436301601 &$1.50_{-0.01}^{+0.01}$  &$75.82_{-2.9}^{+3.1}$&$2.17_{-0.15}^{+0.12}$&$0.22_{-0.07}^{+0.09}$&266.66/323  \\
       P051436301801 &$1.50_{-0.01}^{+0.02}$  &$80.8_{-4.6}^{+2.6}$&$2.05_{-0.19}^{+0.22}$&$0.16_{-0.06}^{+0.12}$&236.00/323  \\
       P051436301802 &$1.51_{-0.01}^{+0.02}$  &$77.5_{-4.5}^{+3.3}$&$2.22_{-0.15}^{+0.17}$&$0.24_{-0.09}^{+0.14}$&283.88/323 \\    
       P051436301901 &$1.51_{-0.01}^{+0.01}$  &$81.0_{-4.0}^{+1.9}$&$2.01_{-0.12}^{+0.20}$&$0.13_{-0.04}^{+0.10}$&269.31/323  \\ 
       P051436302001 &$1.50_{-0.01}^{+0.01}$  &$83.8_{-3.36}^{+2.12}$&$2.01_{-0.20}^{+0.28}$&$0.14_{-0.06}^{+0.09}$&232.00/323 \\ 
       P051436302002
       &$1.53_{-0.01}^{+0.02}$  &$76.2_{-5.9}^{+3.5}$&$2.36_{-0.15}^{+0.20}$&$0.30_{-0.10}^{+0.20}$&239.24/323  \\ 
       P051436302101
       &$1.49_{-0.01}^{+0.01}$  &$93.7_{-3.3}^{+2.3}$&$1.99_{-0.13}^{+0.18}$&$0.17_{-0.05}^{+0.11}$&233.56/323 \\ 
       P051436302102 &$1.48_{-0.01}^{+0.01}$  &$93.8_{-4.5}^{+2.3}$&$1.99_{-0.13}^{+0.18}$&$0.16_{-0.05}^{+0.11}$&254.91/323\\ 
       P051436302103 &$1.48_{-0.01}^{+0.01}$  &$94.3_{-5.9}^{+3.7}$&$2.04_{-0.18}^{+0.20}$&$0.23_{-0.09}^{+0.18}$&265.15/323\\ 
       P051436302201 &$1.45_{-0.01}^{+0.01}$  &$95.6_{-3.3}^{+2.7}$&$1.99_{-0.17}^{+0.18}$&$0.13_{-0.05}^{+0.11}$&244.50/323 \\ 
       P051436302202 &$1.45_{-0.01}^{+0.01}$  &$95.0_{-2.9}^{+2.4}$&$1.76_{-0.20}^{+0.17}$&$0.07_{-0.08}^{+0.03}$&223.11/323 \\ 
       P051436302301 &$1.44_{-0.01}^{+0.01}$  &$93.1_{-3.1}^{+2.1}$&$1.64_{-0.25}^{+0.25}$&$0.06_{-0.03}^{+0.05}$&220.24/323 \\ 
       P051436302302 &$1.43_{-0.01}^{+0.01}$  &$96.3_{-3.2}^{+3.7}$&$1.13_{-0.20}^{+0.39}$&$0.02_{-0.01}^{+0.04}$&246.93/323\\ 
       P051436302401 &$1.43_{-0.01}^{+0.01}$  &$87.3_{-4.3}^{+2.7}$&$1.91_{-0.25}^{+0.24}$&$0.12_{-0.06}^{+0.12}$&203.96/323 \\ 
       P051436302403
       &$1.43_{-0.01}^{+0.01}$  &$89.7_{-4.5}^{+2.4}$&$1.55_{-0.40}^{+0.51}$&$0.03_{-0.02}^{+0.11}$&260.75/323 \\ 
       P051436302404 &$1.43_{-0.01}^{+0.01}$  &$88.3_{-5.4}^{+3.4}$&$2.03_{-0.24}^{+0.26}$&$0.15_{-0.07}^{+0.16}$&218.42/323 \\ 
    \hline
        \label{HXMTpara} &     

    \end{tabular}

\end{table}

\begin{figure*}
\centering

\begin{minipage}[t]{0.45\linewidth}
\centering
\includegraphics[angle=0,scale=0.35]{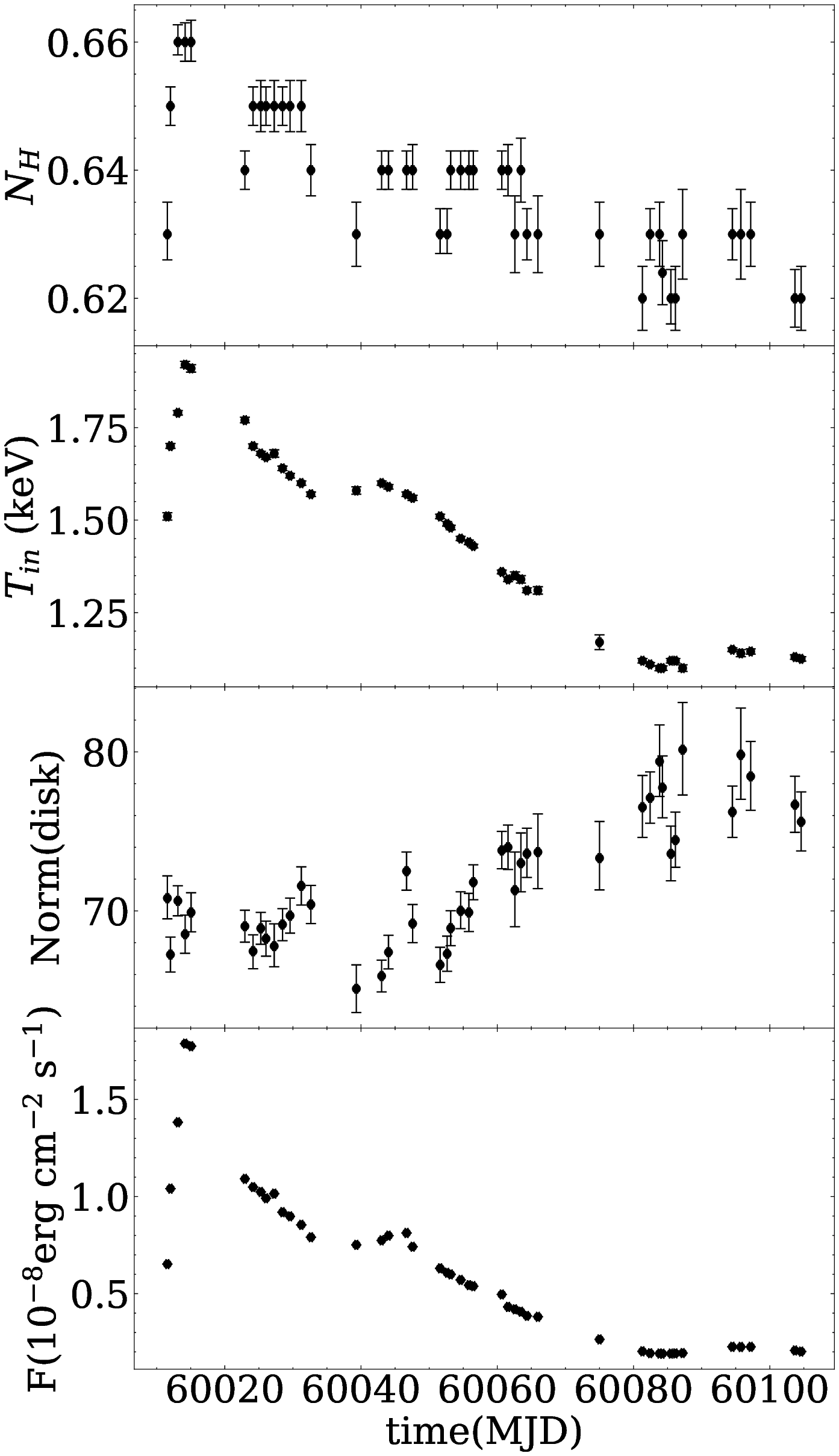}
\end{minipage}%
\hfill
\begin{minipage}[t]{0.45\linewidth}
\centering
\includegraphics[angle=0,scale=0.35]{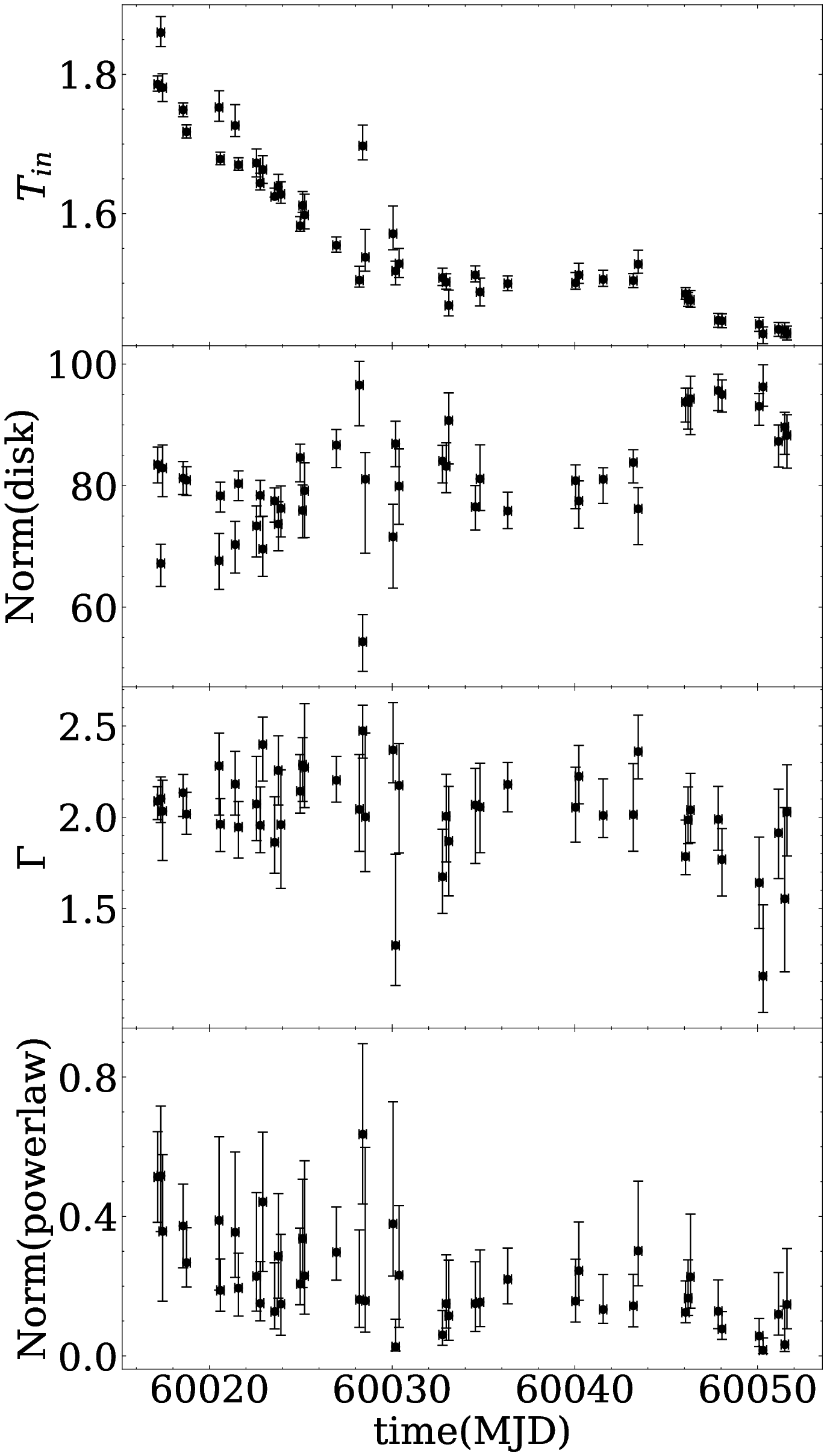}
\end{minipage}
\centering
\caption{Left panel: Evolution of the spectral parameters by the fitting of  \textit{NICER} with M1. $N_{\rm H}$ is the interstellar absorption, $T_{\rm in}$ is the inner disk temperature, Norm(disk) is the normalization of the disk, and F is the flux of the unabsorbed disk from 1 to 100 keV.
Right panel: Evolution of the spectral parameters by the fitting of  \textit{Insigth}-HXMT with M2. $T_{\rm in}$ is the inner disk temperature, Norm(disk) is the normalisation of the disk, $\Gamma$ is the photon index, and Norm(powerlaw) is the normalisation of the powerlaw component. 
}
\label{spec}
\end{figure*}

\begin{figure}
	\centering
	\includegraphics[angle=0,scale=0.35]{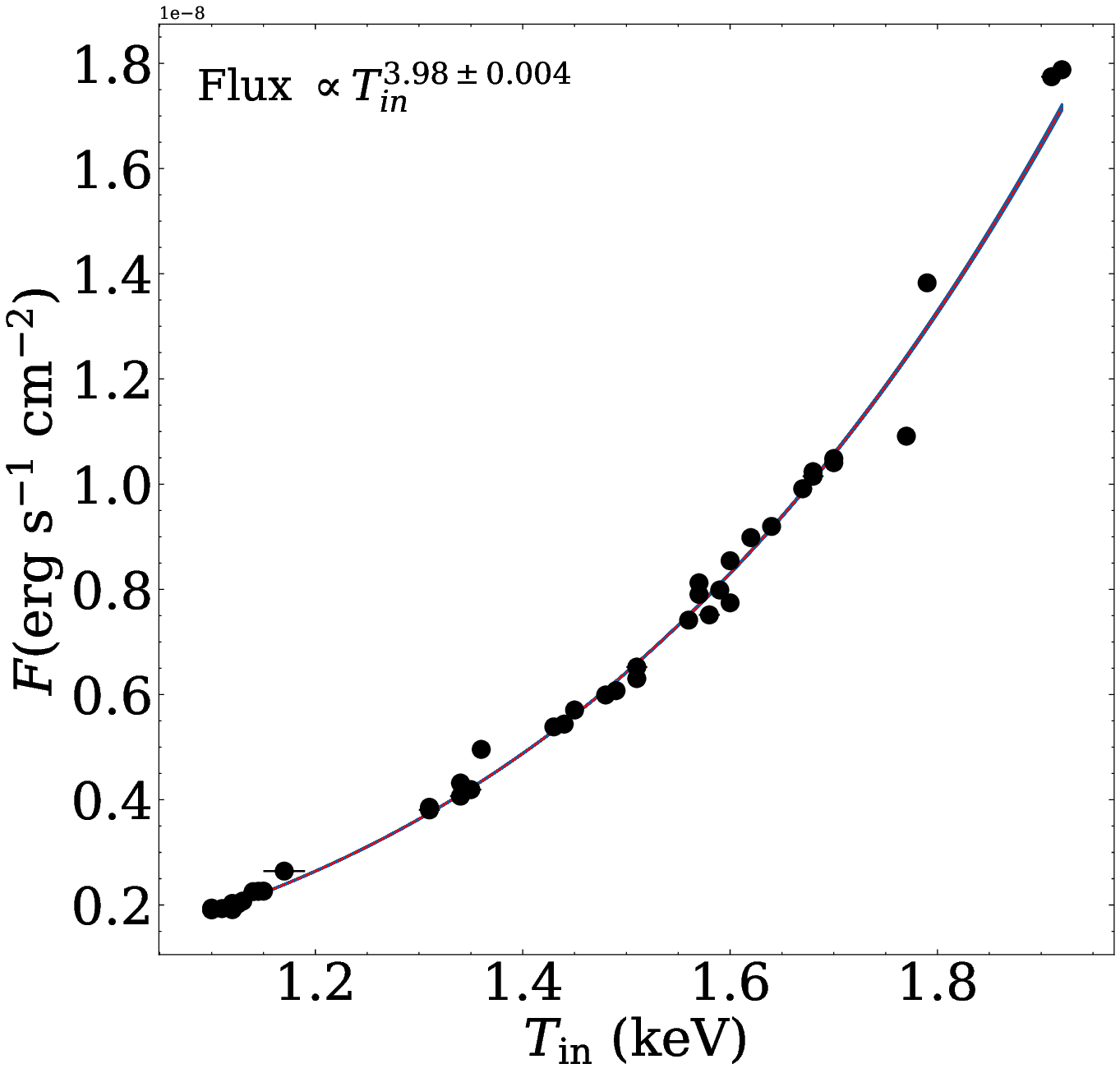}
	\caption{The disk unabsorbed flux (1--100 keV) versus the disk inner temperature ($T_{\rm in}$). Flux is found to correlate with temperature in the form of proportional $T^{3.98}_{\rm in}$}
	\label{T4}
\end{figure}

We use  0.5--10 keV \textit{NICER} data for spectral analysis of SLX 1746--331 during the 2023 outburst.
The first trial for the spectral model (M1) is TBABS*DISKBB, where TBABS accounts for the interstellar absorption \citep{2000Wilms} by considering photoelectric cross-sections provided by \cite{1996Verner} and DISKBB for the multi-temperature blackbody  of the accretion disk \citep{1984Mitsuda}. M1 model results in an acceptable fit, with $\chi^2$/(d.o.f)=142.75/152=0.94.

The flux of the disk in the 1--100 keV is estimated with CFLUX. As shown in the left panel of Figure \ref{spec}, the interstellar absorption remains stable at around $0.64^{+0.02}_{-0.02} \times 10^{22}$ $\rm cm^{-2}$ and the normalization of the disk is around 70, indicating that the inner radius of the disk is relatively stable. The inner disk temperature ($T_{\rm in}$) initially increases and then decreases.
The disk flux versus the inner disk temperature ($T_{\rm in}$) is plotted in Figure \ref{T4}, and a fit with power law results in $\alpha$ around  3.980$\pm$0.004. 
This finding suggests that the disk is at the innermost stable circular orbit (ISCO) during the whole outburst.
We also fit the spectrum of \textit{Insight}-HXMT with M1. Because of the calibration problem of LE, we ignore energies below 2 keV. Accordingly, the column density $ N\rm _H$ can not be well constrained and hence fixed  at 0.64$\times$ $10^{22}$ $\rm cm^{-2}$
and we multiply a CONSTANT component to account for the calibration discrepancies between \textit{Insight}-HXMT LE and ME (in this paper, we fix the constant of LE to 1). At this step, an obvious Compton hump shows up in the residual [$\chi^2$/(d.o.f)=1141.87/325=3.5]. So we add a powerlaw component to fit this residual [$\chi^2$/(d.o.f)=256.11/323=0.79].
Therefore, the alternative fitting model (M2) is CONSTANT*TBABS*(DISKBB+POWERLAW).
The evolution of the spectral parameters is presented in the right panel of Figure \ref{spec}. We find in Figure \ref{spec} that,  while the evolution of the disk component remains almost the same as M1, the normalization of POWERLAW and the photon index $\Gamma$  decrease over time.

\subsection{Properties of the system and the compact object}
\label{estimation}

\begin{figure*}
\centering

\begin{minipage}[t]{0.45\linewidth}
\centering
\includegraphics[angle=0,scale=0.33]{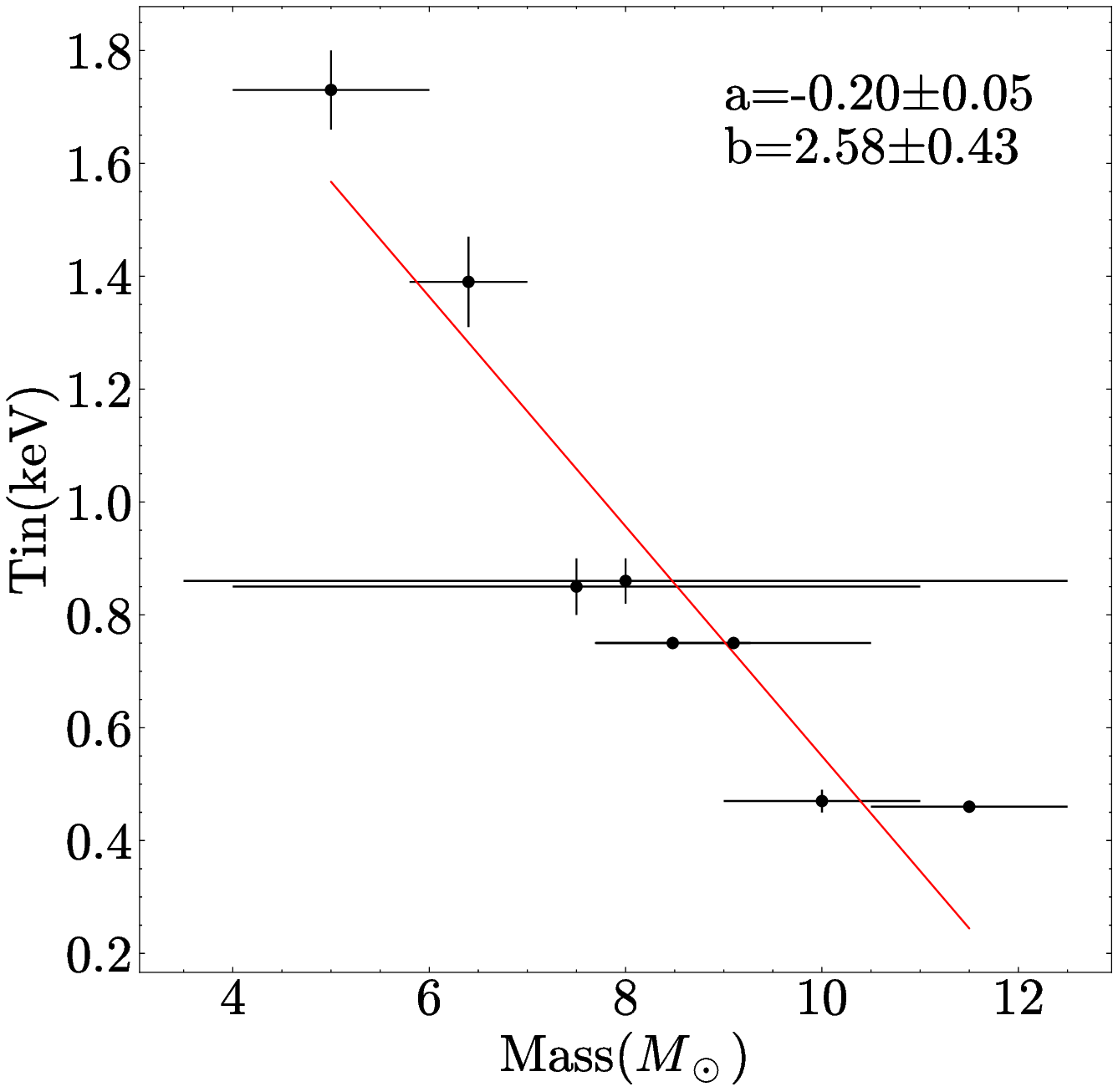}
\end{minipage}%
\hfill
\begin{minipage}[t]{0.45\linewidth}
\centering
\includegraphics[angle=0,scale=0.33]{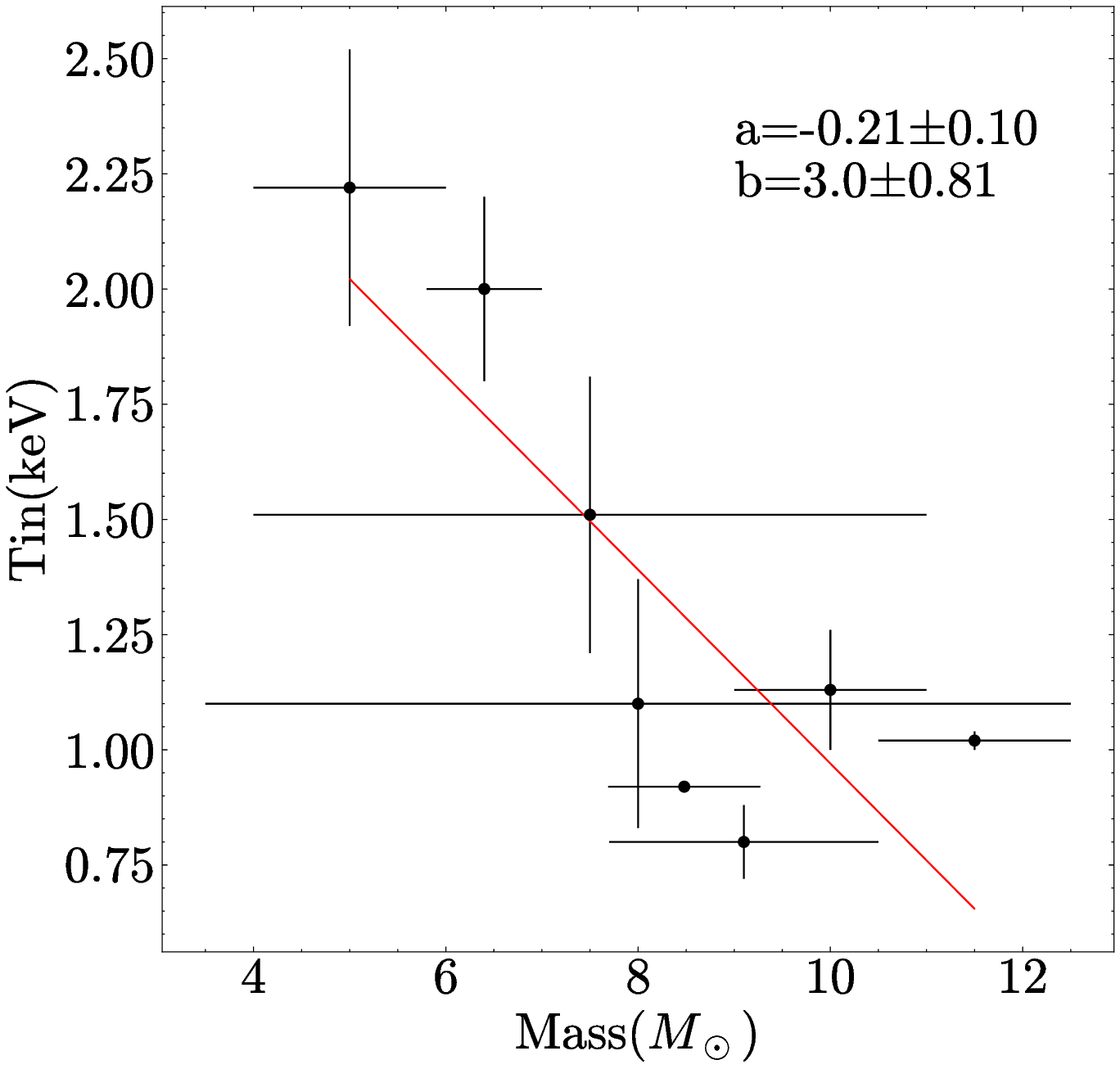}
\end{minipage}
\centering
\caption{The relationship between the mass of a black hole and the maximum temperature of the inner disk in the soft state. Left panel: The observed maximum temperature of the inner disk in the soft state. For different maximum temperatures of disks with different outbursts of BHXRBs, we took the mean value. Right panel: Estimated temperature of the inner disk at a luminosity of 0.3 $L_{\rm Edd} $
}
\label{M-T}
\end{figure*}

\begin{table}[]
    \centering
		\caption{The mass of a black hole X-ray binary and the maximum temperature of its inner disk in the soft state. }
		\begin{threeparttable}
		\begin{tabular}{cccccc}
		 \\ \hline
		   \hline
        \ Sources & Mass & Spin& $T_{\rm in}^{a}$ &$T_{\rm in}^{b}$&References
        \\&$M_\odot$& a &keV &keV &
        \\ \hline
        
        MAXI J1348--630 & $9.1^{+1.6}_{-1.2}$ & $0.78_{-0.04}^{+0.04}$ &$0.75^{+0.01}_{-0.01}$ &$0.80^{+0.08}_{-0.06}$& 1,2,3
        \\ \hline
        
        
         MAXI J1803--298 & 3.5-12.5 & $0.991_{-0.001}^{+0.001}$&$0.86^{+0.04}_{-0.04}$& $1.10_{-0.27}^{+0.27}$&4,5
         \\ \hline
         
        GRS 1758--258 & $\approx 10$ & $0.97^{+0.02}_{-0.05}$&$0.47^{+0.02}_{-0.02}$&$1.13^{+0.13}_{-0.13}$ &6,7,8
         \\ \hline 
         
         MAXI J1727--203 & $ \geq 11.5$ &$0.986^{+0.012}_{-0.159}$& $0.46^{+0.001}_{-0.001}$&$1.02^{+0.02}_{-0.02}$& 9,10
         \\ \hline 
         4U 1957+11 &$5^{+1}_{-1}$ & $0.95^{+0.02}_{-0.04}$&$1.73^{+0.07}_{-0.07}$&$2.22^{+0.3}_{-0.3}$ &10,11,12
         \\ \hline   
         
         V4641 Sgr & $6.4^{+0.6}_{-0.6}$ &$0.86^{+0.02}_{-0.02}$& $1.39^{+0.03}_{-0.13}$&$2.0^{+0.3}_{-0.3}$ &10,13
         
         \\ \hline 
         
         H 1743--322 & $11.21^{+1.65}_{-1.96}$ & $0.98^{+0.01}_{-0.02}$&$0.70^{+0.10}_{-0.14}$&$1.06^{+0.16}_{-0.17}$&10,14,15,16
        \\ \hline
        
         MAXI J1820+070 & $8.48^{+0.79}_{-0.79}$ &$0.988^{+0.006}_{-0.028}$ &$0.75^{+0.01}_{-0.01}$&$0.92^{+0.01}_{-0.01}$& 17,18       \\ \hline
        
        GX 339--4 & 4--11 & $0.95^{+0.02}_{-0.08}$&$0.85^{+0.03}_{-0.12}$&$1.5_{-0.3}^{+0.3}$&19,20
         \\ 
           \hline
    \hline
        \label{mass-a} &     

    \end{tabular}
       \begin{tablenotes} 
       
      \footnotesize
      \item \textbf{NOTE:}
      \item[a] The maximum temperature of the inner disk in the soft state. The average value is taken for those with multiple outbursts.
      \item[b] Estimated inner disk temperatures for 0.3 $L_{\rm Edd}$  \\     \textbf{References.} (1) \cite{2020Jana},(2) \cite{2022Jia}, (3)\cite{2023Dai}, 
     (4) \cite{2022Jana}, (5) \cite{2022Feng},
     (6) \cite{2022Jan}, (7) \cite{2011Soria}, (8) \cite{2001Smith},
     (9) \cite{2022Wang}, (10) \cite{2023Draghis},
     (11) \cite{2021Sharma}, (12)\cite{2008Nowak},
     (13) \cite{2014MacDonald},
     (14) \cite{2017Molla}, (15) \cite{2004CadolleBel}, (16) \cite{2009McClintock},
     (17) \cite{2020Torres}, (18) \cite{2023Peng},
     (19) \cite{2021Shui},(20)  \cite{2019Zdziarski}   
     
      \end{tablenotes}
    \end{threeparttable}
\end{table}

\begin{table}[]
    \centering
		
		\caption{Fitting results with M3, where $D$, $\theta$, and $a$ represent the distance, inclination angle, and spin of the compact object, respectively. The spin of the compact object is derived via M3 fitting under the different parameter sets of $D$, mass, and $\theta$. }
		
		\begin{tabular}{cccc}
        \\ \hline
        \hline
       & $M=3.4M_\odot$ &  $M=5.7M_\odot$ 
        \\
        \hline
        $D$=7.29 kpc & $\theta=70$\textdegree, $a$=$0.80^{+0.01}_{-0.01}$ & $\theta=84$\textdegree, $a$=$0.86^{+0.02}_{-0.04}$ &  \\
        \hline
        $D$=10.81 kpc & $\theta=16$\textdegree, $a$=$0.93^{+0.02}_{-0.03}$ & $\theta=55$\textdegree, $a$=$0.99^{+0.001}_{-0.04}$ &  \\ \hline
        $D=14.33$ kpc & $\theta=0$\textdegree, $a=0.99^{+0.001}_{-0.01}$ & $\theta=18$\textdegree, $a=0.99^{+0.01}_{-0.01}$ & \\
        \hline
        \hline
			\label{spina}
    \end{tabular}
\end{table}

As shown in Figure \ref{T4}, we use M1 to fit the spectrum and obtain a flux of the disk that is proportional to the fourth power of the temperature of the inner disk. For the standard thin disk,
$L_{\rm disk}\approx 4\pi R_{\rm in}^{2}\sigma T_{\rm in}^{4}$, with $R_{\rm in}$, $T_{\rm in}$, and $\sigma$ representing the inner radius, the temperature of the disk and Stefan-Boltzmann constant respectively.
This indicates that the disk is at the ISCO, and the standard disk applies to a luminosity range of 0.001 $L_{\rm Edd}$ $\lesssim $ $L_{\rm disk}$ $\lesssim$0.3 $L_{\rm Edd}$ \citep{2010Steiner,2013Salvesen,2015Garcia,2023Draghis}, where $L_{\rm Edd}$ is the Eddington luminosity ($L_{\rm Edd}=1.26\times 10^{38} \left(\frac{M}{M_\odot}\right)\text{erg/s}$).
As shown in Figure \ref{T4}, we find that the maximum flux for the disk is about ten times greater than the lowest flux.
If we assume that the peak flux has reached a value around 0.3 $L_{\rm Edd}$, we can use the flux and luminosity conversion in X-ray binaries to estimate the mass of the compact object. For a disk blackbody emission, the luminosity of the disk can be approximated as 
$L_{\rm disk} \approx \frac{2\pi F d^2}{\rm cos\theta}$, where the distance $d = 10.81 \pm$3.52 kpc \citep{2015Yan}, and thus a peak luminosity is $L_{\rm disk}=\frac{1.24 \pm 0.81\times 10^{38}} {\rm cos\theta}$ erg/s. By taking  $\theta$ = 0, we can estimate a lower mass limit of the compact object $M \geq$  3.28$\pm 2.14 M_\odot$.

We also take an alternative approach to investigate the possible mass of the compact object hosted in SLX 1736--331. As shown in Table \ref{mass-a}, we constitute a sample by selecting in literature the black hole systems with measured disk temperature in soft state and black mass.  As shown in the left panel of Figure \ref{M-T},  the two parameters of the disk temperature and the black hole mass have an anti-correlation, and a linear fit using Linmix \citep{2007Kelly} takes
an empirical relation of $T_{\rm in} = -0.20_{-0.05}^{+0.05}M + 2.58_{-0.43}^{+0.43}$. For SLX 1746--331, the maximum temperature of the inner disk is found to be $T_{\rm in}$ = 1.9 keV, and according to the above empirical relation, a measurement of disk temperature of 1.9 keV would result in an estimation of the compact object mass of 3.4$\pm 2.3 M_\odot$.  Again, by assuming the outburst peak flux of 0.3 $L_{\rm Edd}$ and a compact mass of 3.3$\pm 2.3 M_\odot$, 
$\rm cos \theta$ is constrained as 0.96$\pm 0.9$. Thus the inclination angle $\theta$ can be constrained as 0--84\textdegree.

For continuum fitting (CF) measurements of spin, it is essential for the accretion disk to extend up to the ISCO \citep{1997Zhang,2005Li}. Typically, the soft state with a disk luminosity $L_{\rm disk} \leq 0.3 L_{\rm Edd}$ is selected for this purpose \citep{1973Shakura,1973Novikov}. 
In order to measure the spin of SLX 1746--331, we replace DISKBB model in M2 with  KERRBB model and convolve it with THCOMP. Therefore, our new model M3 is CONSTANT*TBABS(THCOMP*KERRBB).
The interstellar absorption is set to a fixed value of 0.64 $\times$ $10^{22}$ $\rm cm^{-2}$ based on the \textit{NICER} fit. The electron temperature is also fixed at 150 keV.
As shown in Table \ref{spina}, since the range of inclination angles is relatively large, we choose a few sets for typical parameters of compact star mass, the source distance, and the inclination angle. These parameters are fixed separately in the M3. 
Additionally, the normalization of KERRBB is fixed at 1.0 in accordance with the requirements of the model.
As shown in Table \ref{spina},  all fitting results suggest that SLX 1746--331 hosts a compact object with a relatively high spin value (larger than 0.8). 

For different BHXRBs, the luminosities achieved in each outburst of the soft state are not the same, leading to systematic uncertainties in estimating the empirical relations addressed above.
By assuming a peak flux of 0.3 $L_{\rm Edd}$ for each source, the disk temperatures are revised according to the formula  $L_{\rm disk}\approx 4\pi R_{\rm in}^{2}\sigma T_{\rm in}^{4}$, and shown in Table \ref{mass-a}, a linear fit to the revised disk temperature against the black hole mass results in empirical relations of $T_{\rm in} = -0.21_{-0.1}^{+0.1}M + 2.0_{-0.81}^{+0.81}$. With this updated relationship, the mass of the compact object of  SLX 1746--331 is revised as  $5.2 \pm 4.5M_\odot$. Again,  a low-mass compact is suggested but with a larger uncertainty.

\section{discussion and conclusion}
\label{dis}

We have carried out spectral analysis of the 2023 outburst of SLX 1746--331, which was observed thoroughly by  \textit{Insight}-HXMT and \textit{NICER}. 
Our results reveal for the first time the possible properties of both the outburst and the compact object harbored within this system. The outburst manifested itself as an unusual evolution behavior that stayed mostly in the soft state, with an inner disk remaining around ISCO.  The compact object mass is consistently estimated with different manners: a lower limit of $3.28\pm 2.14 M_\odot$ by assuming 0.3 $L_{\rm Edd}$ of the outburst peak flux, and  $5.2 \pm 4.5M_\odot$ from the empirical correlation between disk temperature and black hole mass,  established from a sample of black hole systems.

The previous observations on SLX 1746--331 with RXTE/PCA and INTEGRAL/JEM-X have already pointed to a transition nature and its soft spectrum, which is very similar to the soft state of a black hole, and thus it was identified as a black hole transient X-ray binary \citep{1990Skinner,1996White,2003Homan}. 
The distance was estimated by \cite{2015Yan} to be 10.81$\pm 3.52$ kpc using RXTE data, but the basic parameters of black hole mass and inclination of the system are still missing.
When fitting the empirical relation, uncertainties are propagated to the final measurement of the black hole mass, resulting in a wider range of black hole mass. However, by incorporating the black hole mass obtained from the range of Eddington luminosities, we can place a constraint on the black hole mass of SLX 1746--331.
Obviously, given the current outburst behavior of  SLX 1746--331, a more constrained lower limit on the compact object can be derived if the source distance could be better measured in the future, and the mass can be directly estimated if an inclination angle could be available.

With the thorough observations of \textit{Insight}-HXMT and \textit{NICER}, the outburst of SLX 1746--331 is found with rather peculiar evolution behavior. The outburst was dominated by soft emission and the inner disk remained at ISCO during the almost entire outburst. Such behavior is rather different from the normal outburst experienced by the outburst of the black hole systems, which usually have a low hard state and intermediate state. Although rare, a similar outburst was also found in another black hole system MAXI J0637--430. Like SLX 1746--331, MAXI J0637–-430 shows an outburst that lacks the hard state during the rising phase and directly transitions to the soft state \citep{2022Ma}. Therefore, SLX 1746--331 and MAXI J0637--430 may constitute a special sample of the black hole X-ray binary system with less prominent hard and intermediate states in their outbursts.

A comparison between SLX 1746--331 and MAXI J0637--430 shows the possible common feature in the mass of their compact object. \cite{2022Soria} estimate the black hole mass of MAXI J0637--430 to be about 5.1$\pm1.6 M_\odot$, while the mass of the compact object (most likely a black hole) hosted in SLX1746--331 is measured around  $5.2 \pm 4.5M_\odot$. However, these two systems are distinct in other aspects:  MAXI J0637--430 hosts 
a low-spin (less than 0.25) black hole, underwent an outburst of less than 0.1 $L_{\rm Edd}$ peak luminosity and a maximum disk temperature of 0.7 keV; while  SLX 1746--331 harbors a compact object with the relatively high spin of larger than 0.8 and is observed with a disk temperature of as high as 1.9 keV. 
MAXI J0637--430 has a donor star with a mass of 0.25$\pm 0.07 M_\odot$ and the shortest binary period $P_{\rm orb} \approx 2.2^{+0.8}_{-0.6}$ hr \citep{2022Ma,2022Soria}, while for SLX 1746--331 the mass of the donor star and the orbital period are yet unknown. 
Both systems underwent outbursts without experiencing a clear hard state, which means that the transition from hard to soft states, if any, should have occurred at low accretion rates. State transition from hard to soft states has been addressed in \cite{2021Cao}, where they think the critical accretion rate for state transition from hard to soft could be highly related to the large-scale magnetic field born out of the disk.  The transition is balanced between the timescales of ion-electron equilibration and accretion of the ADAF. Note that the dependence of the disk magnetic field is positive for the former and negative for the latter, a transition to the soft state under a lower accretion rate in principle needs a low disk magnetic field. Since the disk is formed from matter accreted from the companion via Roche-Lobe, a system with a small compact object and tight orbital period could in principle result in a smaller accretion disk, and hence a weaker large-scale disk magnetic field. 
Therefore, the similar outburst behavior of SLX 1746--331 and MAXI J0637-430 might suggest both sources have small orbits and light compact objects, which are already confirmed for MAXI J0637--430. Furthermore, the compact object mass as measured in these two systems most likely falls into the long-debated mass gap of 3--5$M_\odot$ of the black holes,  which makes them interesting observational targets for future observations by e.g.,  \textit{Einstein Probe (EP)} \citep{2018Yuan}. Capturing the initial rising phase of the outbursts by \textit{EP} would tell us if for such systems the low hard state really exists, as well the luminosity level for transition to the soft state.


\begin{acknowledgments}
This work is supported by the National Key R\&D Program of China (2021YFA0718500), the National Natural Science Foundation of China under grants U1838201, U1838202,  U2038101, U1938103, 12273030, U1938107.
This work made use of data and software from the \textit{Insight}-HXMT mission, a project funded by China National Space Administration (CNSA) and the Chinese Academy of Sciences(CAS). This work was partially supported by International Partnership Program of Chinese Academy of Sciences (Grant No.113111KYSB20190020).
This research has made use of  software provided by of data obtained from the High Energy Astrophysics Science Archive Research Center (HEASARC), provided by NASA’s Goddard Space Flight Center.
L. D. Kong is grateful for the financial support provided by the Sino-German (CSC-DAAD) Postdoc Scholarship Program (57251553).

\end{acknowledgments}

\bibliography{sample631}{}
\bibliographystyle{aasjournal}



\end{document}